# Multi-Point Detection of the Powerful Gamma Ray Burst GRB221009A Propagation through the Heliosphere on October 9, 2022.

Andrii Voshchepynets,[1,2] Oleksiy Agapitov,[2,3] Lynn Wilson III,[4] Vassilis Angelopoulos,[5] Samer T. Alnussirat,[2] Michael Balikhin,[6] Myroslava Hlebena,[1] Ihor Korol,[7,8] Davin Larson,[2] David Mitchell,[2] Christopher Owen,[9] and Ali Rahmati[2]

[1]*Department of System Analysis and Optimization Theory,*
*Uzhhorod National University,*
*Uzhhorod, Ukraine*
[2]*Space Sciences Laboratory:*
*University of California Berkeley*
*Berkeley, CA 94720*
[3]*Astronomy and Space Physics Department, National Taras Shevchenko University of Kyiv, Kyiv, Ukraine*
[4]*Goddard Space Flight Center*
*National Aeronautics and Space Administration*
*Greenbelt, MD*
[5]*Department of Earth, Planetary, and Space Sciences,*
*University of California Los Angeles*
*Los Angeles, CA*
[6]*University of Sheffield,*
*Sheffield, UK*
[7]*Department of Algebra and Differential Equation,*
*Uzhhorod National University,*
*Uzhhorod, Ukraine*
[8]*Department of Mathematical Analysis,*
*The John Paul II Catholic University of Lublin,*
*Lublin, Poland*
[9]*Mullard Space Science Laboratory:*
*University College London*
*Dorking RH5 6NT, UK*

## ABSTRACT

We present the results of processing the effects of the powerful Gamma Ray Burst GRB221009A captured by the charged particle detectors (electrostatic analyzers and solid-state detectors) onboard spacecraft at different points in the heliosphere on October 9, 2022. To follow the GRB221009A propagation through the heliosphere we used the electron and proton flux measurements from solar missions Solar Orbiter and STEREO-A; Earth's magnetosphere and the solar wind missions THEMIS and *Wind*; meteorological satellites POES15, POES19, MetOp3; and MAVEN - a NASA mission orbiting Mars. GRB221009A had a structure of four bursts: less intense Pulse 1 - the triggering impulse - was detected by gamma-ray observatories at $T_0$=13:16:59 UT (near the Earth); the most intense Pulses 2 and 3 were detected on board all the space-craft from the list, and Pulse 4 detected in more than 500 s after Pulse 1. Due to their different scientific objectives, the spacecraft, which data was used in this study, were separated by more than 1 AU (Solar Orbiter and MAVEN). This enabled tracking GRB221009A as it was propagating across the heliosphere. STEREO-A was the first to register Pulse 2 and 3 of the GRB, almost 100 seconds before their detection by spacecraft in the vicinity of Earth. MAVEN detected GRB221009A Pulses 2, 3, and 4 at the orbit of Mars about 237 seconds after their detection near Earth. By processing the time-delays observed we show that the source location of the GRB221009A was at RA 288.5°, Dec 18.5° ±2° (J2000).

## 1. INTRODUCTION

GRB 221009A was a bright and long-lasting gamma-ray burst (GRB) detected by the space-borne telescopes such as Burst Alarm Telescope (BAT) on board Swift (Dichiara et al. 2022) (afterglow), GBM and LAT on board the FGST (Veres et al. 2022; Lesage et al. 2022; Pillera et al. 2022), AGILE (Piano et al. 2022; Ursi et al. 2022), INTEGRAL SPI-ACS (Gotz et al. 2022), Solar Orbiter (Xiao et al. 2022), SRG/ART-XC (Lapshov et al. 2022), Konus-WIND (Frederiks et al. 2022), GRBAlpha (Ripa et al. 2022), and STPSat-6 (Mitchell et al. 2022), High Energy Burst Searcher (HEBS) on SATech-01 (Liu et al. 2022), BepiColombo (Kozyrev et al. 2022) and by the ground observatories such as LHAASO (The Large High Altitude Air Shower Observatory) (Brdar & Li 2023). The GRB 221009A started with a precursor at $T_0$ = 2022-10-09 13:16:59 UTC (Pulse 1), followed by a set of pulses at ∼ $T_0$+225s (Pulse 2), ∼ $T_0$+256s (Pulse 3) and ∼ $T_0$+509s (Pulse 4) (GRB Coordinates Network, GCN (Veres et al. 2022; Liu et al. 2022; Rodi & Ubertini 2023)). The afterglow of the burst outshone all other GRBs seen before (Sahu et al. 2023) in a very high energy (VHE) band. The water Cherenkov detector array (WCDA) and the larger air shower kilometer square area (KM2A) detector at LHAASO observed more than 5000 VHE photons in the 500 GeV–18 TeV energy range within 2000 s from the trigger, making them the most energetic photons ever observed from a GRB (Baktash et al. 2022). The multi-wavelength afterglow and comparative brightness of the GRB221009A were studied in Laskar et al. (2023) and Kann et al. (2023). The event was so long and intense that the increased ionization by X- and γ-ray emission caused sudden global ionospheric disturbances in the D-region of Earth's ionosphere (Hayes & Gallagher 2022; Pal et al. 2023).

The source location of the GRB has been derived to be centered at RA=288.282° and Dec=19.495° (J2000) with a 90% containment radius of 0.027° (Pillera et al. 2022) and its redshift estimated to be z=0.15 (Veres et al. 2022; de Ugarte Postigo et al. 2022). The relatively small value of the redshift indicates that this is one of the closest observed long-duration GRBs (Rastinejad et al. 2022; Troja et al. 2022; Mei et al. 2022; Gompertz et al. 2023) as its luminosity distance is ≈ 720 Mpc (O'Connor et al. 2023) based on ΛCDM cosmological model (Freedman 2021). Even after taking into account the proximity, the GRB 221009A remains one of the most luminous explosions to date (O'Connor et al. 2023), with lower limits of isotropic-equivalent gamma-ray energy $\gtrsim 3 \times 10^{54}$ erg measured over the energy range from 20 keV to 10 MeV (Frederiks et al. 2022; de Ugarte Postigo et al. 2022; Kann

& Agui Fernandez 2022) and more precise value of $\gtrsim 1.2 \times 10^{55}$ (Frederiks et al. 2023; Lesage et al. 2023).

In this article, we present the analysis of observations of GRB 221009A by the charged particle detectors onboard various spacecraft located at different points across the heliosphere on October 9, 2022. Data from charged particle detectors on board heliosphere spacecraft can provide an additional perspective on the GRB that complements observations from the gamma-ray telescopes (Schwartz et al. 2005; Terasawa et al. 2005). Charged particle detectors, despite not being designed to detect high-energy gamma-ray photons, can measure the flux of secondary particles produced within the material of the spacecraft (Pisacane 2005) or the detector itself (Battiston et al. 2023). GRB 221009A has been previously studied using data from the electrostatic analyzer and solid state telescopes onboard THEMIS mission (Agapitov et al. 2023), the HEPP-L charged particle detector on board the low Earth orbit (LEO) China Seismo-Electromagnetic Satellite (Battiston et al. 2023) and MEPED onboard POES/MetOp(Vitale et al. 2023). It was shown that the high-time resolution of the electron and proton flux measurements (up to 8 measurements per second (Agapitov et al. 2023)) reveal the structure of the intense bursts (Pulses 2, 3, and 4) and were shown to follow quite well the fine inner structure of the Pulses 2 and 3 of the GRB221009A signal recorded by HEBS with 0.05 s resolution (Liu et al. 2022). Here we present observations of the GRB221009A by the charged particle detectors of STEREO-A, Solar Orbiter (SolO), *Wind*, THEMIS, POES/MetOp, and MAVEN spacecraft. STEREO-A (Kaiser et al. 2008) and SolO (Müller et al. 2020) are Sun observing missions. *Wind* (Wilson III et al. 2021) and THEMIS (Sibeck & Angelopoulos 2008) are the missions dedicated to studying processes in the solar wind and Earth's magnetosphere. POES (Rodger et al. 2010) and MetOp (Selesnick et al. 2020) are a constellation of polar-orbiting weather satellites. MAVEN (Jakosky et al. 2015) is the NASA mission to study atmospheric losses at Mars. The spacecraft locations are shown in Figure 1. The description of the missions' particle flux measurements is in Section 2. The multi-point observations of the GRB221009A effects and the results of timing processing are in Sections 3 and 4. The conclusion section contains a short description of the obtained results.

## 2. DATA DESCRIPTION

### 2.1. *SolO Data*

The Solar Orbiter mission (SolO) is a joint ESA and NASA project, which aims to study how the Sun creates and controls the heliosphere (Müller et al. 2020; Owen et al. 2020). SolO has a comprehensive set of remote-sensing and *in-situ* probing instruments that enable correlative studies of the solar wind plasma – its origin, transport processes, and elemental composition. Energetic Particle Detector (EPD) (Rodríguez-Pacheco et al. 2020) onboard SolO is composed of four units: the SupraThermal Electrons and Protons (STEP), the Electron Proton Telescope (EPT), the Suprathermal Ion Spectrograph (SIS), and the High-Energy Telescope (HET). For this study, we used data provided by the EPT. The EPT consists of two double-ended telescopes: EPT 1 is pointing sunward and anti-sunward along the nominal Parker spiral while EPT-HET 2 is pointing northward and southward. The EPT relies on the magnet/foil technique to separate and measure electrons and protons in the energy range [0.03-0.43] MeV for electrons and [0.05-5.83] MeV for protons. EPT sensors are able to collect data with a cadence of up to 1 second.

### 2.2. *STEREO A Data*

The twin STEREO spacecraft, A and B, were launched on October 26, 2006. The purpose of the mission is to understand the causes and mechanisms of coronal mass ejection (Kaiser et al. 2008). As of now, only STEREO-A is still operational. In this study, we used data gathered by the IMPACT (In situ Measurements of Particles And CME Transients) suite of instruments, which measures the flux of solar wind electrons, energetic electrons, protons, and heavier ions (Luhmann et al. 2008). The Suprathermal Electron (STE) instrument, part of the IMPACT, is designed to measure the $\sim$2–100 keV suprathermal electrons in the interplanetary medium near 1 AU STE utilizes silicon semiconductor detectors (SSDs) that measure all energies simultaneously, and has a temporal resolution of 10 seconds.

### 2.3. *Wind Data*

*Wind* spacecraft is a part of the Global Geospace Science program (GGS), a subset of the International Solar-Terrestrial Physics (ISTP) program, designed to study basic plasma processes occurring in the near-Earth solar wind (Acuña et al. 1995). *Wind* currently orbits at the first Lagrange point, about $\sim 200\ R_E$ sun-ward from Earth. It carries a comprehensive set of instruments, including a gamma-ray spectrometer KONUS (Aptekar et al. 1995) and a particle experiment 3DP (Lin et al. 1995). KONUS is a gamma-ray spectrometer designed primarily for the detection and study of cosmic gamma-ray bursts, with capabilities for the study of solar flares in the hard X-ray region (Wilson III et al. 2021; Lysenko et al. 2022). The 3DP (Three-Dimensional Plasma and energetic particle investigation) consists of three detector systems: the semi-conductor detector telescopes (SST), the electron electrostatic analyzers (EESA), and the ion electrostatic analyzers (PESA). The SST consists of three arrays, each with a pair of double-ended telescopes to measure electron and ion fluxes above 20 keV. EESA and PESA are electrostatic analyzers that measure particle fluxes in a wide range of energies from 3 eV to 30 keV.

### 2.4. *THEMIS Data*

THEMIS (Time History of Events and Macroscale Interactions during Substorms) was designed to carry out multi-point investigations of substorm phenomena in the tail of the terrestrial magnetosphere (Sibeck & Angelopoulos 2008). It consists of five identically equipped satellites, which carry ElectroStatic Analyzers (ESA) and Solid-State Telescopes (SST) particle detectors (McFadden et al. 2008). THEMIS ESA is a pair of back-to-back top hat hemispherical electrostatic analyzers that measure the distribution functions of ions (0.005 to 25 keV) and electrons (0.005 to 30 keV) over $4\pi$ str at 22.5°x22.5° angular resolution, $\Delta E/E$ 30% energy resolution and 3s temporal resolution (4 s for the two ARTEMIS spacecraft (Angelopoulos 2014) orbiting the Moon (the Acceleration, Reconnection, Turbulence, and Electrodynamics of the Moon's Interaction with the Sun mission)) time resolution plasma moments (McFadden et al. 2008). SST measures the superthermal (0.03 - 1MeV) part of the ion and electron distributions over $4\pi$ str with similar energy, angular and temporal resolution. Combining the energy and solid-angle measurements can sufficiently improve time resolution (up to a sub-second cadence (Agapitov et al. 2023)).

### 2.5. *POES and MetOp Data*

POES (Polar Orbiting Environmental Satellites) and MetOp (Meteorological Operational Satellites) are a series of polar-orbiting meteorological satellites operated by NOAA (National Oceanic and Atmospheric Administration) and EUMETSAT (European Organisation for the Exploitation of

| sc | SolO | STEREO A | THEMIS | | *Wind* | | POES/MetOp | MAVEN | | |
|---|---|---|---|---|---|---|---|---|---|---|
| det | EPT | STE | ESA | SST | PESA | SST | MEPED | SWEA/SWIA | SEP | STATIC |
| e$^-$ | 30−420 | 20−100 | 5e-3−30 | 30−1e3 | | 25−400 | 30−2.5e3 | 3e-3−30 | 20−1e3 | |
| i$^+$ | 50−8.8e3 | | 5e-3−25 | 0.03−1e3 | 3e-3−30 | 0.02−6e3 | 0.03−7e3 | 5e-3−25 | 0.02−6e3 | 1e-3−30 |

**Table 1.** Summary of the instruments used in this study (the energies are in keV)

Meteorological Satellites) on LEO orbits (Rodger et al. 2010; Selesnick et al. 2020). The current spacecraft generation - POES 15, 19, and three MetOp satellites are equipped with the same instruments MEPED (Medium Energy Proton and Electron Detector), which is a subsystem of Space Environment Monitor (SEM-2) (Pettit et al. 2021). MEPED is a set of solid-state energetic particle detectors that monitor the intensities of proton and electron flux over an energy range extending from 30 keV to ∼ 7 MeV, for protons, and up to 2.5 MeV, for electrons, with 2 s cadence.

### 2.6. *MAVEN Data*

MAVEN (The Mars Atmosphere and Volatile EvolutioN) is a NASA mission dedicated to studying the ionosphere and upper atmosphere of Mars, its interactions with the Sun, and the solar wind (Jakosky et al. 2015). The spacecraft carries a comprehensive set of charged particle measurement instruments designed to study losses of the ionospheric plasma due to interaction with the solar wind. It consists of the SolarWind Electron Analyzer (SWEA), the Solar Wind Ion Analyzer (SWIA), the Solar Energetic Particle (SEP) instrument, and the Supra-Thermal And Thermal Ion Composition (STATIC) instrument. SWEA is a top hat electrostatic analyzer utilizing electrostatic deflectors to provide a field of view of 360°×120° and measures electrons between energies of 3 eV and 5 keV with 2 s cadence (Mitchell et al. 2016). The SWIA is an electrostatic analyzer that measures the energy and angular distributions of solar wind and magnetosheath ions with a 4-s cadence (Halekas et al. 2015). It uses deflection optics, that provide a broad 360°×90° field of view and has a broad energy range of 5 eV to 25 keV. SEP consists of two sensors, each utilizing a dual, double-ended solid-state detector telescope that measures fluxes of electrons from 20 to 1000 keV and ions from 20–6000 keV (Larson et al. 2015). SEP measurements have a 2 s time cadence. STATIC is an electrostatic top hat analyzer designed to measure the cold ion of the Martian ionosphere, the heated suprathermal tail of the plasma in the upper ionosphere, and the pickup ions (McFadden et al. 2015). It operates over an energy range of 0.1 eV up to 30 keV, with a base time resolution of 4 seconds. STATIC utilizes electrostatic deflectors that provide a field of view of 360°×90°.

Table 1 summarizes the parameters of the instruments used in this publication.

### 3. OBSERVATIONS

Figure 1 shows the locations of the spacecraft during the GRB221009A. Panel (a) provides a general view of the spacecraft's location in the Heliocentric coordinate (HCI) system (Fränz & Harper 2002). Relative to the Sun, SolO was the closest, at the heliocentric distance of ∼0.30 AU (Astronomical unit ∼ 499 light-second), followed by STEREO-A (∼0.96 AU), Earth, and MAVEN (∼1.46 AU). During the event, SolO was below the ecliptic plane (shown as a blue surface). Panels (b) and (c) show a zoomed view of the configuration of the spacecraft in the vicinity of the Earth in the Geocentric Solar Ecliptic (GSE) coordinate system (Fränz & Harper 2002). The THEMIS mission (Sibeck & Angelopoulos 2008) consists of five identically equipped satellites (probes THA, THB, THC, THD, and THE). In 2011, two of the probes (THB and THC) were repurposed for the ARTEMIS mission and moved first to a Lissajous orbit around a lunar Lagrange point and a year later into highly eccentric, near-equatorial lunar orbit (Angelopoulos 2014). During the event, THB and THC were located near the Moon in the tail of Earth's magnetosphere at geocentric distances of 58 $R_E$ and 61 $R_E$ ($R_E$ - Earth radius $\sim$ 0.021 light-second), while the inner THEMIS probes were at the geocentric distances of $\sim$9-10 $R_E$. *Wind* was further from the Earth to the sunward direction (*x*-axis) at the geocentric distance of $\sim$203.26 $R_E$ from the Earth. POES/MetOp spacecraft are on polar LEO orbits (panel (c) in Figure 1). The propagation of the GRB 221009A in the heliosphere is schematically illustrated by the set of planar surfaces in Figure 1.

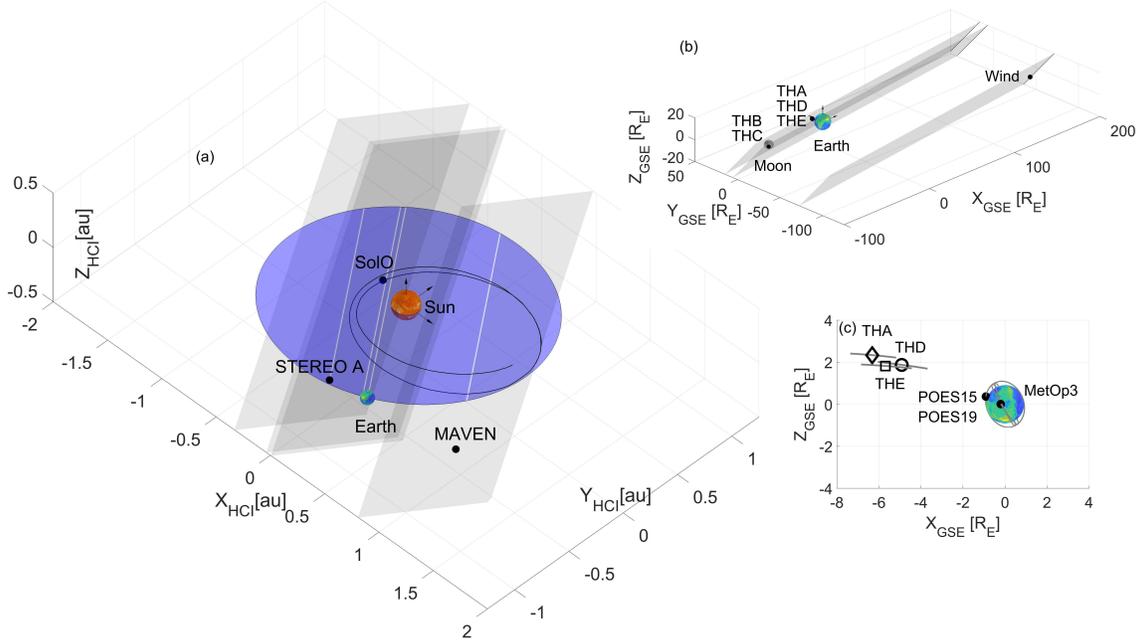

**Figure 1.** The configuration of the spacecraft during the GRB 211211A event: panel (a) - location of STEREO A, SolO, MAVEN in HCI (Heliocentric inertial) coordinate system. The Earth's orbit is shown in blue. Panels (b) and (c) display the locations of THEMIS and POES/MetOp spacecraft in the Geocentric Solar Ecliptic (GSE) coordinate system. The propagation of the GRB 221009A across the heliosphere is schematically shown with the set of planes.

(Xiao et al. 2022) reported the detection of GRB221009A by the Spectrometer/Telescope for Imaging X-rays (STIX) (Krucker et al. 2020) onboard SolO (panel (a) in Figure 2). In addition, Pulses 2,3, and 4 were captured by the EPT electron and proton detectors (they are shown in Figures 2b and 2c respectively). The electron and proton apparent flux enhancements associated with GRB221009A are observed at energies up to 0.5 MeV. Figures 2e and Figures 2f show the integrated electron and proton counts in EPT 2 and the anti-sunward detector on EPT 1. For comparison, the shifted by $-13$ s temporal profile measured by SST on THEMIS is shown in gray. The temporal profiles recorded by EPD follow quite well the profiles recorded on THEMIS (Agapitov et al. 2023) and HEPP-L (Battiston et al. 2023) but EPD detected GRB221009A $\sim$ 13 s earlier: the start of the main peak of Pulse 2 at $\sim T_0+212$s, and the start of Pulse 3 at $\sim T_0+243$s, compared to $T_0+225$s and $T_0+256$s reported in (Agapitov et al. 2023) ($T_0 = $ 13:16:59.000 (Veres et al. 2022)). Data from STE onboard STEREO-A is shown in Figure 2d. GRB221009A-related signals were detected in the energy range from 20 keV to 100 keV at 13:19:06 (Pulse 2) and 13:19:36 (Pulse 3) almost 100 seconds earlier than they were detected by THEMIS (Agapitov et al. 2023). The STE time revolution of 10 seconds, is not sufficient to resolve details like shoulders (Liu et al. 2022) on the temporal profile of the burst (Figure 2g).

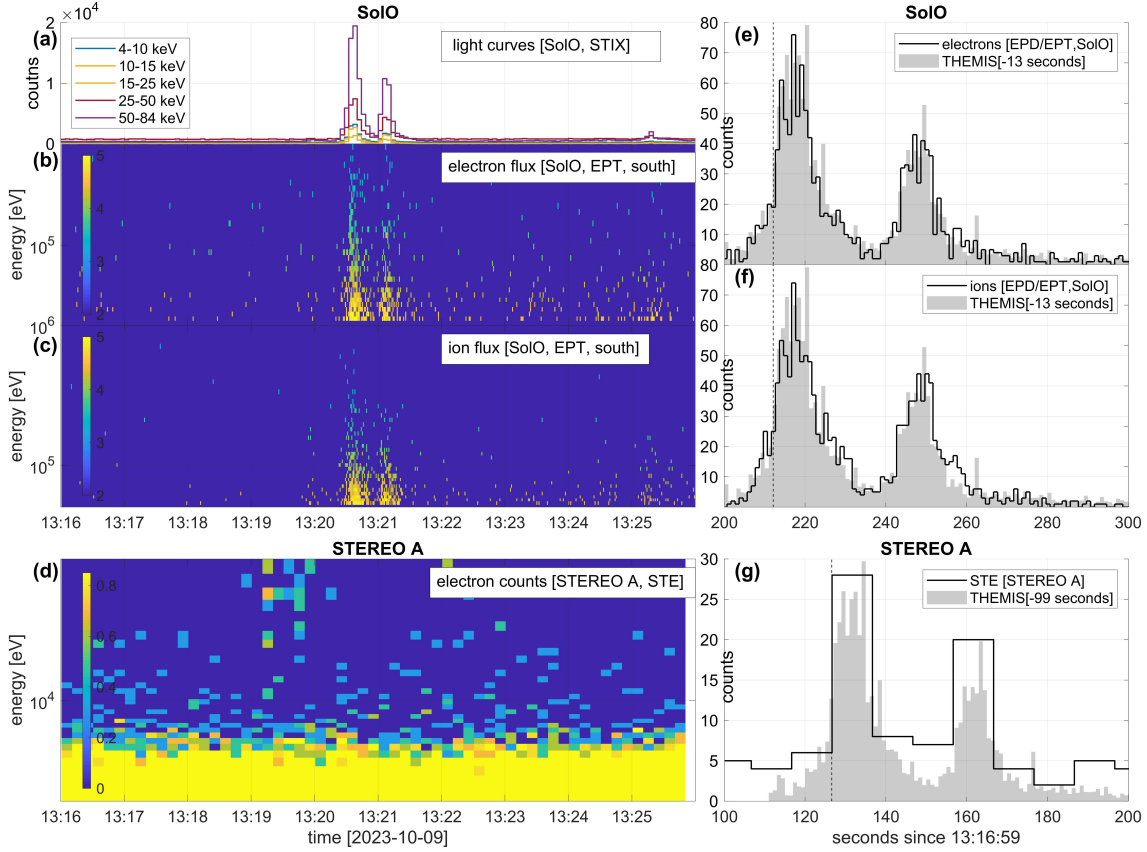

**Figure 2.** Detection of the effects of the GRB221009A by SolO and STEREO-A: (a) - light-curves detected by STIX (SolO); (b) and (c) - electron and ion flux from EPT (SolO); (d) - electron counts from STE (STEREO-A); (e) and (f) - integrated electron and ion counts from EPT (SolO); (g) - integrated electron counts from STE (STEREO-A). The background light-gray silhouette is the light curves of the combined THC and THC SST proton flux counts from (Agapitov et al. 2023).

KONUS onboard *Wind* detected the first Pulse of the GRB221009A at 13:17:01.648 (Frederiks et al. 2022) (also shown in Figure 3a). Figure 3b and 3c show electron and ion flux measured by the *Wind* SST with 12 s resolution. The GRB221009A-related signals (Pulses 2 and 3) are visible in the lower energy channels: 25-40 keV for electrons and 70-130 keV for ions. The related signal was also found in the $\alpha$-particle density calculated onboard from the PESA measurements. Comparing the temporal profile with the THEMIS measurements (Figure 3d) shows a time delay of $\sim 1.5$s.

The GRB221009A was detected by THEMIS ESA and SST on borad the four THEMIS probes (THA, THB, THC, and THE). The signal appears as two very intense bursts at 13:20:36-13:21:30 and a subsequent less intense burst at around 13:25:30 (Figure 3e). The event was studied in detail by (Agapitov et al. 2023). It was shown that the temporal profile of the burst with 0.25 s resolution can be obtained by combining the spin-resolved data from THB and THC Agapitov et al. (2023). Here we use the light curve of the combined THB and THC SST counts with 1 s resolution for comparison with the profiles detected by the other spacecraft used in this study.

Figure 3f shows electron flux measured by the POES 19 MEPED detector. The effects of the GRB221009A are visible as two relatively intense bursts detected within an energy range of 40 keV - 1 MeV at 13:20:30-13:21:30. The temporal profiles of the GRB221009A-related signals recorded on POES 15,19 and MetOp are shown in panel (g). The profile of Pulse 2 is very similar to that detected by THC, while the profile of Pulse 3 is much less prominent. The event was studied in detail by (Vitale et al. 2023).

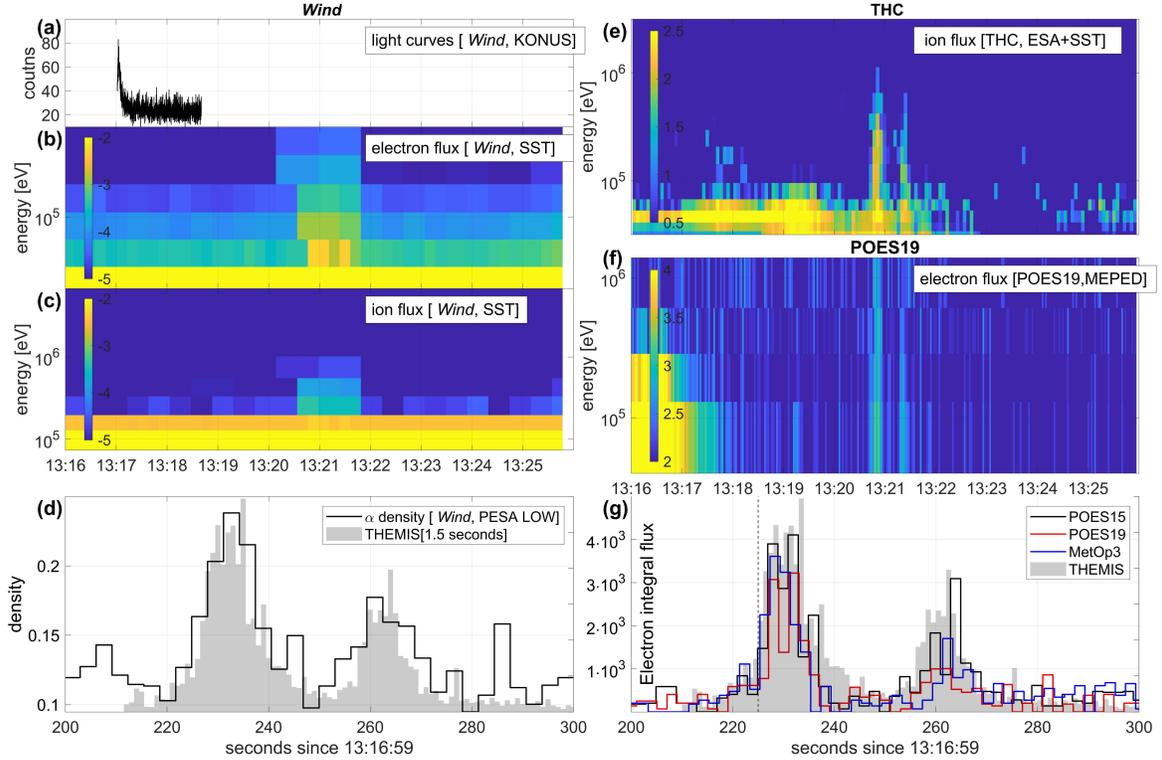

**Figure 3.** Detection of the effects of the GRB 221009A by the spacecraft of Earth group: (a) - light curve by KONUS (*Wind*, data available at https://gcn.gsfc.nasa.gov/konus_grbs.html); (b) and (c) - electron and ion flux by SST (*Wind*); (d) - density of $\alpha$-particles by PESA (*Wind*); (e) - ions flux by ESA+SST (THC); (f) - electron flux by MEPED (POES19); and (g) -the integrated electron flux by MEPED (POES15, POES19, and MEtOp).

Figure 4 shows the observations of the GRB221009A made by the particle instruments onboard MAVEN. The two very intense bursts - Pulses 2 and 3 - were detected by all of the MAVEN particle instruments between 13:24:30 and 13:25:30, almost 4 minutes later than they were detected in the vicinity of the Earth (Earth-Mars distance at that time was ∼0.73 AU or ∼364 light second). STATIC (Figure 4d) detected the third burst - Pulse 4 - at 13:29:30. Measurements by the electrostatic analyzers are consistent, and the GRB221009A-related signal is visible in the ion data (SWIA and STATIC) through the entire energy range (up to 30 keV). In the electron measurements by SWEA, the signal was detected at the energies above ∼ 1 keV. Electron measurements by SEP show the effects of the gamma-ray burst within the energy range from 20 to 200 keV. Comparison of the temporal profiles measured by MAVEN with the THEMIS observation (Figure 4e-h) shows a time delay of about 237.5s.

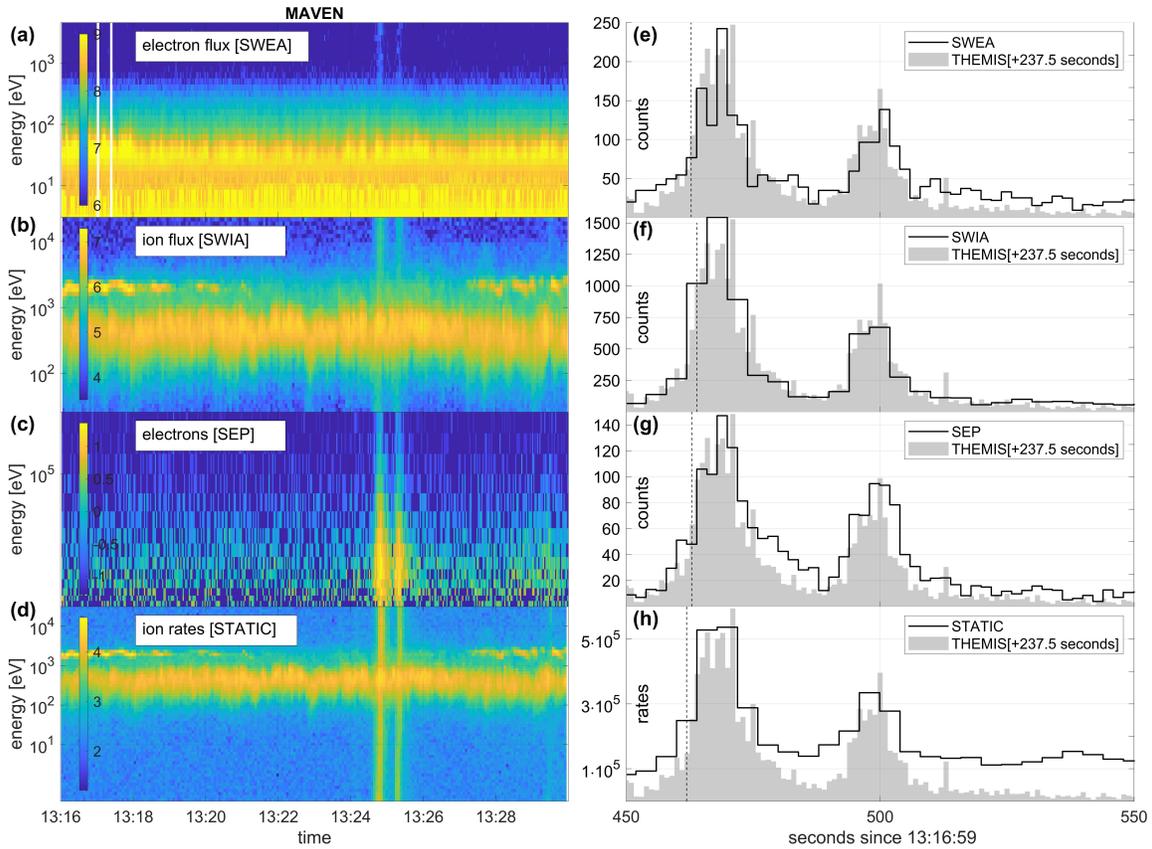

**Figure 4.** Detection of the effects of the GRB221009A by MAVEN: (a) - electron flux by SWEA; (b) ion flux by SWIA; (c) electron counts by SEP; (d) ion rates by STATIC; (e)-(h) the integrated counts/rates from (a)-(d).

High-energy gamma-ray photons mainly interact with the spacecraft and detector material via the electron-positron pair production process. The flux and energies of these secondaries are determined by the initial photon's energy and flux as well as the thickness and density of the passive material of the spacecraft and detector itself. For this GRB, the spacecraft particle detectors mainly encountered energy deposition from the secondary charged particles (electrons and positrons). The simulation performed for the parameters of the HEPP-L telescope (Battiston et al. 2023) showed that the 83% of the detected secondaries originate in the passive material around the silicon detector and about 17% were directly generated within the detector. The secondary electrons also produce the signal recorded by the proton telescopes (and electrostatic analyzers (Agapitov et al. 2023)), as was shown by the comparison of the electron fluxes measured by the electron and proton MEPED telescopes on board POES/MetOp (Vitale et al. 2023). The dilated study of the signal detected by SolO and MAVEN will be a subject of the upcoming publications.

4. TRACKING PROPAGATION OF THE GRB221009A ACROSS THE HELIOSPHERE

The multi-point observations of the GRB221009A signal propagation by the spacecraft at different locations enable estimation of the normal to the signal front (Paschmann & Daly 1998) and thus can be used to determine the direction to its source in the sky. Using THC spacecraft as a reference and assuming that the burst propagated as a planar front, one can find a normal **n** to the front by solving a system of linear equations:

$$\sum_{\beta=1}^{3} \left(r_\beta^\alpha - r_\beta^{THC}\right) n_\beta = v \left(T^\alpha - T^{THC}\right),$$

where **r** is a position vector of the spacecraft, $T$ is a time of observation, $v$ is the speed of the front, and $\alpha =$ is the spacecraft: STEREO, SolO, and MAVEN. Positions of the spacecraft in the HCI coordinate system are listed in the legend of Figure 5. Using the start of the Pulse 2 detected at $T_0$ +126±5s at STEREO-A, +212±0.5s at SolO, +225±0.5s at THEMIS, and +462.5±1s at MAVEN, we estimated normal vector **n**=[-0.6033, -0.3775, 0.7025] and the $v = 1.01c$. The normal vector gives the GRB221009A source location at RA 288.5°, Dec 18.5° in J2000 a cone of uncertainty (95% confidence level) $\delta\theta = 2°$ . Start times of the Pulse 2 were determined by comparing the temporal profiles with the profiles captured on THC (Agapitov et al. 2023) (shown as gray dashed lines in Figures 2-4). In the case of MAVEN, the start of the Pulse 2 was found to be at $T_0$+262s (STATIC) and $T_0$+263s (SWEA and SEP). In order to estimate the accuracy of the obtained values, we used the approach described in Vogt et al. (2011), which is based on the reciprocal vector formalism and takes into account uncertainties both in the crossing times and in the spacecraft positions.[1] [2]

The obtained results are consistent with the previously reported values of RA = 288.282°, Dec = 19.495° ±0.027° (Pillera et al. 2022), RA=288.2643°, Dec=19.7712°(Dichiara et al. 2022) and RA=287.761°, Dec=20.670° (center) (Kozyrev et al. 2022).

Figure 5 summarizes the observations of GRB221009A by the spacecraft across the heliosphere sorted along the normal to the GRB221009A signal front (the $y$-axis) on October 9, 2022. It shows the temporal profiles of the gamma-ray burst signal recorded on STEREO A, SolO, THEMIS, *Wind*, and MAVEN, starting at $T_0$ =13:16:59.000 - time of the detection of the triggering signal (Pulse 1) of the gamma-ray burst near the Earth (Veres et al. 2022). All of the listed spacecraft have detected the effects of Pulses 2 and 3, seen as two intense bursts. SolO, THEMIS, and MAVEN have also captured Pulse 4. The speed-of-light signal propagation is marked by the light-gray lines.

## 5. CONCLUSION

We report the effects of the powerful Gamma Ray Burst GRB221009A recorded by the charged particle detectors aboard the spacecraft probes at different locations across the heliosphere: Sun and solar wind observatories - STEREO A, Solar Orbiter, and *Wind*; the magnetosphere multi-probe mission THEMIS; space weather probes at LEO orbits - POES/MetOp; and the planetary Martian mission MAVEN. GRB221009A had a structure of four bursts: the first triggering impulse was detected by the gamma-ray observatories near the Earth at $T_0$=13:16:59 UT, the most intense Pulses 2, 3, and Pulse 4 were detected in the charged particle detector records (Pulse 4 - in more than 500 s after the triggering Pulse 1). The effects of the gamma-ray burst were detected at times corresponding to the spacecraft locations: STEREO-A detected the signal of Pulse 2 almost 100 seconds before it was detected in the vicinity of the Earth, and it took another ∼237 seconds for the signal to reach MAVEN. Assuming that the burst propagated as a planar front across the heliosphere, we have determined the direction towards the source region of the GRB221009A at RA 288.5°, Dec 18.5° with a 95% containment radius of 2°, for the first time, from multi-point observations of spacecraft-based particle detectors.

---

[1] The biggest source of the uncertainties is the STEREO time resolution of $\Delta T$=10s. To estimate the influence of the spacecraft array geometry accuracy, we analyzed a position tensor in terms of its eigenvectors and the tetrahedron geometric parameters: planarity $P$, elongation $E$, and the RMS inter-spacecraft distance $L$. For the configuration of STEREO, SolO, THC, and MAVEN, $\delta\theta = v\Delta T/(L(1-E)) \sim 2°$ and $\delta v/V \sim 2\%$.

[2] In the case of more than four spacecraft available and/or the pre-known propagation speed (the speed of light), the least squares method could be applied to solve the system (Paschmann & Daly 1998). For the GRB221009A case, adding data from *Wind* spacecraft results in values RA=288.4°, Dec=18.9°. The method could be used in the planar configuration if data is available from only three spacecraft. In this case, two components of the front normal can be found from the system of equations on the plane. An absolute value of the third component of the normal could be found if the speed of the front is known. Using data from THC, STEREO, and MAVEN in the $(x-y)$ plane one can find **n**=[-0.5998,-0.4424, ±0.5047]. If a positive $z$ component is taken, the location of the burst is RA=292°, Dec=8°. This method is less accurate than the one described above.

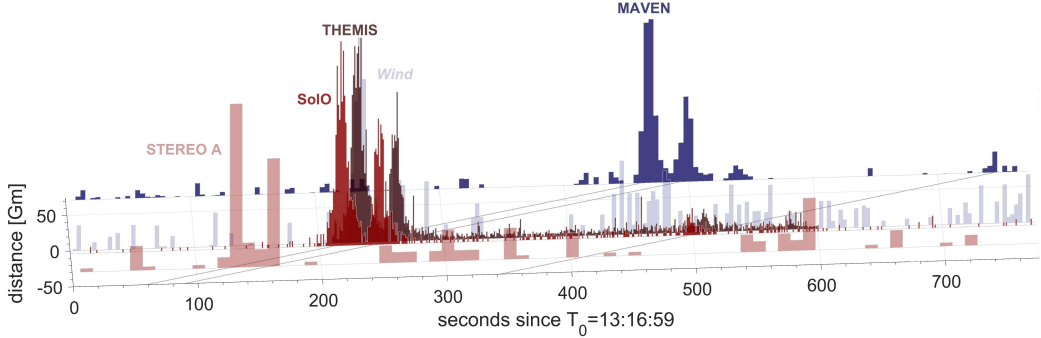

**Figure 5.** Propagation of the GRB221009A across the heliosphere. The temporal profiles of the gamma-ray burst signal detected by the charged particle detectors onboard STEREO-A, SolO, THC, *Wind*, and MAVEN are shown with different colors. $T_0 = 13{:}16{:}59$ - time of the detection of the triggering impulse of GRB221009A in the vicinity of the Earth. $Y$-axis - the distance between the points of observations along the normal to the signal front, where 0 is the position of the Earth. The normal was estimated based on the coordinates and times provided in the legend. Light-gray lines illustrate the propagation of Pulses 2,3, and 4 of the GRB221009A with the speed of light.

We acknowledge NASA contract NAS5-02099 for the use of data from the THEMIS Mission and J. P. McFadden for the use of THEMIS ESA data and MAVEN STATIC data; J. Luhman for the use of STEREO-A data; Samuel Krucker for the use of SolO STIX data. The work was supported by NSF grant number 1914670, NASA's Living with a Star (LWS) program (contract 80NSSC20K0218), and NASA grants contracts 80NNSC19K0848, 80NSSC22K0433, 80NSSC22K0522, 80NSSC20K0697, and 80NSSC20K0697. STEREO-A data processed at SSL under NASA grant 80NSSC22K1359.